\documentclass[10pt,twocolumn,amsmath,amssymb,aps,prb,showpacs,longbibliography,superscriptaddress]{revtex4-1}
\usepackage[latin9]{inputenc}
\usepackage{textcomp}
\usepackage{mathrsfs}
\usepackage{amsmath}
\usepackage{amssymb}
\usepackage{graphicx}
\usepackage{esint}

\makeatletter

\usepackage{amsfonts}
\usepackage{graphics}

\makeatother

\begin{document}
\title{Unconventional superconductivity of an altermagnetic metal: Polarized
BCS and inhomogeneous Fulde-Ferrell-Larkin-Ovchinnikov states}
\author{Hui Hu}
\affiliation{Centre for Quantum Technology Theory, Swinburne University of Technology,
Melbourne 3122, Australia}
\author{Zhao Liu}
\affiliation{Centre for Quantum Technology Theory, Swinburne University of Technology,
Melbourne 3122, Australia}
\author{Xia-Ji Liu}
\affiliation{Centre for Quantum Technology Theory, Swinburne University of Technology,
Melbourne 3122, Australia}
\date{\today}
\begin{abstract}
We investigate the superconductivity of two-dimensional spin-1/2 Fermi
systems with $d$-wave altermagnetism under external magnetic field
near zero temperature. At large altermagnetic coupling without magnetic
field, we show that altermagnetism drives a second-order phase transition
from the standard Bardeen-Cooper-Schrieffer (BCS) state to an inhomogeneous
Fulde-Ferrell-Larkin-Ovchinnikov (FFLO) state. The inclusion of magnetic
field turns the BCS state into a long-sought polarized BCS superconductor
with spin-population imbalance. It also shrinks the parameter window
of the FFLO state and eventually leads to a nontrivial quantum tri-critical
Lifshitz point, where two second-order phase transition lines between
the polarized BCS, FFLO and normal states intersect. At small altermagnetic
coupling, we find the usual route to the FFLO state driven by magnetic
field. The presence of the altermagnetic coupling narrows the phase
window of the FFLO state and creates another quantum Lifshitz point,
where a first-order transition curve meets a second-order transition
line. Between the two Lifshitz points, the transition from the polarized
BCS state to the normal state is smooth. Our predicted rich phase
diagram is relevant to some recently discovered unconventional magnets,
including RuO$_{2}$ that exhibits a relatively high superconducting
temperature in the thin film limit under applied strain. Our results
of unconventional superfluidity are also testable in ultracold atom
laboratories, where a spin-1/2 altermagnetic Fermi gas might be realizable
upon loading into two-dimensional Hubbard lattices.
\end{abstract}
\maketitle

\section{Introduction}

At absolute zero temperature, the possibility of superconductivity
or superfluidity in spin-population imbalanced interacting Fermi systems
remains elusive \citep{Casalbuoni2004,Radzihovsky2010,Gubbels2013,Kawamura2024}.
The best known theoretical proposal is the so-called Fulde-Ferrell-Larkin-Ovchinnikov
(FFLO) state driven by an external magnetic field, as independently
suggested by Fulde and Ferrell (FF) \citep{Fulde1964} and by Larkin
and Ovchinnikov (LO) \citep{Larkin1964} in 1960s. In the FFLO state,
Cooper pairs carry a nonzero center-of-mass momentum due to the mismatched
Fermi surfaces for each spin component. As a result, the pairing order
parameter becomes spatially inhomogeneous. However, despite enormous
efforts in theory and experiment over the past six decades \citep{Ohashi2002,Uji2006,Hu2006,Hu2007,Liu2007,Kenzelmann2008,Liao2010,Liu2013,Cao2014,Gerber2014,Takahashi2014,Sheehy2015,Wang2018,Kawamura2022,Wan2023,Zhao2023},
there is so far no smoking-gun evidence that such an exotic finite-momentum
superconductor has been observed. The second best known candidate
of imbalanced superconductivity is probably the Sarma state \citep{Sarma1963}
or the breached pairing state \citep{Liu2003}, which condensates
at zero momentum. This polarized, BCS like state typically corresponds
to a local maximum in the energy landscape, and therefore requires
some very specified conditions to thermodynamically stabilize \citep{Forbes2005,Zou2018}.

Most recently, in a pioneering study \citep{Chakraborty2024} Chakraborty
and Black-Schaffer surprisingly suggested that a FFLO state can be
induced by $d$-wave altermagnetism in the absence of magnetic field
in the newly discovered altermagnetic metals \citep{Hayami2019,Uchida2020,Hayami2020a,Hayami2020b,Smejkal2020,Smejkal2022,Mazin2022,Mazin2023,Hiraishi2024,Amin2024}.
Although there is no net magnetism due to the unconventional, momentum-dependent
form of the collinear magnetic order \citep{Smejkal2022}, the FFLO
state was predicted to arise in the presence of an attractive $d$-wave
interaction potential \citep{Chakraborty2024}. This interesting finding
was attributed to the $d$-wave symmetry of the superconducting order
parameter whose gapless nodes coincide with the altermagnetic nodes.
The FFLO state disappears for an $s$-wave interaction potential \citep{Chakraborty2024}.
On the other hand, the altermagnetism induced FFLO state was found
to smoothly connect the magnetic field induced FFLO state \citep{Chakraborty2024}.
Therefore, the transition from the BCS state to the FFLO state, induced
by either altermagnetism or magnetic field, is first order. However,
in a subsequent research by Hong, Park and Kim \citep{Hong2025},
it was clearly demonstrated that an $s$-wave on-site interaction
can accommodate the FFLO state at a fixed chemical potential, through
the formation of Bogoliubov Fermi surface. Moreover, the transition
from the BCS state to the altermagnetism induced FFO state is smooth
\citep{Hong2025}. To reconcile the two controversial observations,
on the stabilization mechanism of the FFLO state and on the order
of the phase transition, clearly we need more in-depth exploration
of the altermagnetism induced FFLO state. 

We note that, the finite-momentum Cooper pairing \citep{Zhang2024}
and gapless superconducting state \citep{Wei2024} in proximitized
altermagnets has also been predicted in a number of interesting theoretical
studies. Unconventional superconductivity in altermagnetic metals
with spin-orbit coupling has been addressed \citep{Zhu2023,deCarvalho2024}.
Josephson effect \citep{Ouassou2023,Lu2024,Maeda2024,Fukaya2025}
and Andreev reflection \citep{Papaj2023,Sun2023} in altermagnets
have been discussed.

In this work, we would like to examine the existence of the FFLO state
in a $d$-wave altermagnetic metal with attractive $s$-wave interaction
at a given particle density. This fixed particle density removes a
slight but potentially important difference in the setups considered
in the previous two studies, if we seriously take into account the
very fragile nature of the FFLO state. 

\begin{figure}
\begin{centering}
\includegraphics[width=0.5\textwidth]{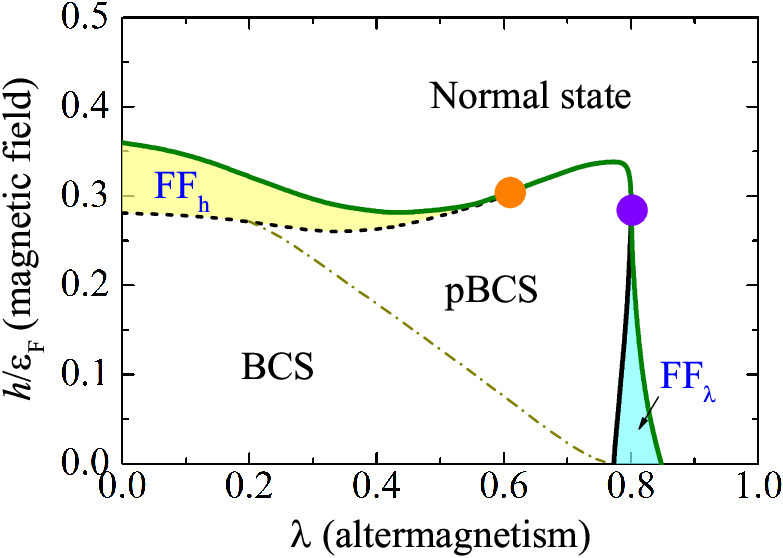}
\par\end{centering}
\caption{\label{fig1: phasediagram} The phase diagram at the binding energy
$\varepsilon_{B}=0.08\varepsilon_{F}$ and at a negligible temperature
$T=0.01T_{F}$. Two different FF phases, mainly driven by magnetic
field (i.e., the $\textrm{FF}_{h}$ phase) and by altermagnetism (see
the $\textrm{FF}_{\lambda}$ phase), are shown by the yellow and blue
areas, respectively. The transition from the BCS phase to the $\textrm{FF}_{h}$
phase is first order, as denoted by the dashed line. All the other
transitions are second order and are shown by solid lines. The two
dots indicate the two quantum tri-critical Lifshitz points, where
the BCS, FF and normal states meet together. The dot-dashed line divides
the BCS phase into two parts, a spin-population balanced BCS state
(with imbalance $\delta n=(n_{\uparrow}-n_{\downarrow})<0.001n$)
and a spin-population partially polarized BCS state. The latter is
abbreviated as the pBCS state.}
\end{figure}

Our results are briefly summarized by a rich phase diagram in Fig.
1, in the plane of the altermagnetic coupling strength (i.e., $\lambda$
in the horizontal axis) and the magnetic field ($h$ in the vertical
axis). On the one hand, at large altermagnetic coupling we confirm
the altermagnetism induced FFLO state in the presence of attractive
$s$-wave interaction, as observed by Hong, Park and Kim without magnetic
field \citep{Hong2025}. We also confirm the smooth phase transitions
from the BCS state to the FFLO state, and eventually to the normal
state. Our extended study with the inclusion of magnetic field further
reveals a non-trivial quantum tri-critical Lifshitz point \citep{Hornreich1975,Gubbels2009,Pisarski2019},
where two second-order phase transition curves intersect. On the other
hand, at small altermagnetic coupling we find the usual magnetic field
induced FFLO state \citep{Sheehy2015} and the transition from the
BCS state to the FFLO is first order. This inevitably leads to another
quantum Lifshitz point, where a first-order phase transition line
meets a second-order phase transition line. The existence of two quantum
Lifshitz points clearly suggests that the FFLO states induced by magnetic
field and altermagnetism are fundamentally different. 

Remarkably, in the phase diagram we also find a large phase space
for a polarized BCS superconductivity near the absolute zero temperature.
This unexpected polarized BCS state might be understood as the reminiscence
of the long-sought Sarma phase \citep{Sarma1963}, whose thermodynamic
instability is efficiently cured by the altermagnetism. The direct
transition from the polarized BCS state to the normal state is smooth
and locates between the two quantum Lifshitz points.

The rest of the paper is organized as follows. In the next chapter
(Sec. II), we outline the model Hamiltonian that describes a $d$-wave
altermagnetic metal in external magnetic field. In Sec. III, we briefly
discuss how to solve the model Hamiltonian by using the standard mean-field
theory, with an inhomogeneous FF order parameter. In Sec. IV, we present
some details of our numerical calculations, which could be useful
to improve the numerical accuracy. In Sec. V, we discuss in detail
the two distinct kinds of the FF states driven by external magnetic
field and altermagnetism, respectively. We show the existence of a
polarized BCS phase due to the combined effect of magnetic field and
altermagnetism. We also determine the nature of phase transitions
between different states. Finally, we conclude in Sec. VI and present
some outlooks for future works.

\section{Model Hamiltonian}

Let us start from the following model Hamiltonian for a two-dimensional
Fermi system with the $d$-wave altermagnetic term \citep{Smejkal2022}
$J_{\mathbf{k}}\equiv\lambda\hbar^{2}k_{x}k_{y}/(2m)$ and external
magnetic field $h$ along $z$-direction, $\mathscr{H}=\int d{\bf x}({\cal H}_{0}+{\cal H}_{int})$,
where the Hamiltonian densities are,
\begin{eqnarray}
{\cal H}_{0} & = & \sum_{\sigma=\uparrow,\downarrow}\psi_{\sigma}^{\dagger}\left({\bf x}\right)\left[\hat{\xi}_{{\bf k}}+s(\sigma)\left(\hat{J}_{\mathbf{k}}+h\right)\right]\psi_{\sigma}\left({\bf x}\right),\\
{\cal H}_{int} & = & U_{0}\psi_{\uparrow}^{\dagger}\left({\bf x}\right)\psi_{\downarrow}^{\dagger}\left({\bf x}\right)\psi_{\downarrow}\left({\bf x}\right)\psi_{\uparrow}\left({\bf x}\right).
\end{eqnarray}
Here, we have defined the operators $\hat{\xi}_{{\bf k}}\equiv-\hbar^{2}{\bf \nabla}^{2}/(2m)-\mu$,
$\hat{k}_{x}=-i\partial_{x}$, $\hat{k}_{y}=-i\partial_{y}$ and $\hat{J}_{\mathbf{k}}\equiv\lambda\hbar^{2}\hat{k}_{x}\hat{k}_{y}/(2m)$,
and have denoted $s(\uparrow)=+1$ and $s(\downarrow)=-1$. The dimensionless
parameter $\lambda$ characterizes the strength of altermagnetism.
The chemical potential $\mu$ in $\hat{\xi}_{{\bf k}}$ can be tuned
to yield the given total particle density $n=n_{\uparrow}+n_{\downarrow}$.
The Fermi system may or may not have a nonzero spin-population imbalance
$\delta n=n_{\uparrow}-n_{\downarrow}$, depending on its ground state
in the phase diagram. 

For the inter-particle interaction, we consider a short-range $s$-wave
contact potential. In two dimensions, its interaction strength $U_{0}$
can be conveniently regularized by using a two-body binding energy
$\varepsilon_{B}$ \citep{He2015}:
\begin{equation}
\frac{1}{U_{0}}=-\frac{1}{\mathcal{S}}\sum_{\mathbf{k}}\frac{1}{\hbar^{2}k^{2}/m+\varepsilon_{B}},
\end{equation}
where $\mathcal{S}$ is the area of the Fermi system.

We note that, our model Hamiltonian provides an excellent description
of some recently discovered unconventional magnets, particularly RuO$_{2}$
\citep{Uchida2020,Hiraishi2024}, although its classification as an
altermagnet is still under debate. The model Hamiltonian is also potentially
within reach in cold-atom laboratories \citep{Das2024}. By loading
a spin-1/2 Fermi gas of fermionic atoms into a specifically designed
two-dimensional lattice, $d$-wave altermagnetism may arise due to
the Hubbard on-site repulsion \citep{Das2024}. In addition, the external
magnetic field can be introduced by tuning the imbalance between the
two spin populations and an effective short-range attraction between
fermions can be mediated by doping some bosonic atoms \citep{DeSalvo02019}.

\section{Mean-field theory}

To search for the FF superconductivity, we assume a FF-like order
parameter $\Delta({\bf x})=-U_{0}\left\langle \psi_{\downarrow}\left({\bf x}\right)\psi_{\uparrow}\left({\bf x}\right)\right\rangle =\Delta\exp[i\mathbf{q}\cdot\mathbf{x}]$
and consider the standard mean-field decoupling of the interaction
Hamiltonian density \citep{Hu2006,Liu2013}, 
\begin{equation}
{\cal H}_{int}\simeq-\left[\Delta({\bf x})\psi_{\uparrow}^{\dagger}\left({\bf x}\right)\psi_{\downarrow}^{\dagger}\left({\bf x}\right)+\text{H.c.}\right]-\frac{1}{U_{0}}\left|\Delta({\bf x})\right|^{2}.
\end{equation}
The direction of the FF momentum $\mathbf{q}$ is determined by the
altermagnetism. It can be either along the $x$-axis or along the
$y$-axis, due to the $C_{4z}\mathcal{T}$ symmetry of the altermagnetic
term \citep{Hong2025}. Here, we choose the $x$-axis for concreteness
and set $\mathbf{q}=q\mathbf{e}_{x}$ and $\Delta(\mathbf{x})=\Delta\exp[iqx]$.
In the framework of the mean-field Bogoliubov-de Gennes (BdG) theory,
the total Hamiltonian can be rewritten into the form \citep{Hu2006,Liu2013},
\begin{equation}
{\cal \mathscr{H}}_{MF}=\int d{\bf x}\Phi^{\dagger}\left({\bf x}\right){\cal H}_{BdG}\Phi\left({\bf x}\right)-\frac{\Delta^{2}}{U_{0}}\mathcal{S}+\mathcal{E}_{0},
\end{equation}
where $\Phi\left({\bf x}\right)\equiv[\psi_{\uparrow}\left({\bf x}\right),\psi_{\downarrow}^{\dagger}\left({\bf x}\right)]^{T}$
is the Nambu spinor and 
\begin{equation}
{\cal H}_{BdG}\equiv\left[\begin{array}{cc}
\hat{\xi}_{{\bf k}}+\hat{J}_{\mathbf{k}}+h & -\Delta\left({\bf x}\right)\\
-\Delta^{*}\left({\bf x}\right) & -\hat{\xi}_{{\bf k}}+\hat{J}_{\mathbf{k}}+h
\end{array}\right],\label{eq:HBdG}
\end{equation}
and $\mathcal{E}_{0}'=\sum_{\mathbf{k}}(\hat{\xi}_{{\bf k}}-\hat{J}_{\mathbf{k}}-h)$
is an energy shift (to be specified below), due to the abnormal order
in the field operators $\psi_{\downarrow}(\mathbf{x})$ and $\psi_{\downarrow}^{\dagger}({\bf x})$
taken in the Nambu spinor representation.

\subsection{Bogoliubov quasiparticles}

As the FF order parameter takes a plane-wave form, it is convenient
to diagonalize the mean-field Hamiltonian in momentum space, with
the following Bogoliubov equations \citep{Liu2013}, 
\begin{equation}
{\cal H}_{BdG}\Phi_{{\bf k}}\left({\bf x}\right)=E_{{\bf k}}\Phi_{{\bf k}}\left({\bf x}\right),
\end{equation}
where 
\begin{equation}
\Phi_{{\bf k}}\left({\bf x}\right)\equiv\left[\begin{array}{c}
u_{{\bf k}\uparrow}e^{+iqx/2}\\
v_{{\bf k}\downarrow}e^{-iqx/2}
\end{array}\right]e^{i{\bf kx}}
\end{equation}
and $E_{{\bf k}}$ are the wavefunction and energy of the Bogoliubov
quasiparticles, respectively. With this wavefunction, the energy shift
is $\mathcal{E}_{0}'=\sum_{\mathbf{k}}(\xi_{{\bf k}-\mathbf{q}/2}-J_{\mathbf{k}-\mathbf{q}/2}-h)$,
where $\xi_{\mathbf{k}}\equiv\hbar^{2}k^{2}/(2m)-\mu$, and the Bogoliubov
equations recast into the form,
\begin{equation}
\left[{\cal H}_{BdG}\right]\left[\begin{array}{c}
u_{{\bf k}\uparrow}\\
v_{{\bf k}\downarrow}
\end{array}\right]=E_{{\bf k}}\left[\begin{array}{c}
u_{{\bf k}\uparrow}\\
v_{{\bf k}\downarrow}
\end{array}\right],
\end{equation}
where $\left[{\cal H}_{BdG}\right]$ is a 2 by 2 matrix given by,
\begin{equation}
\left[\begin{array}{cc}
\xi_{{\bf k}+\mathbf{q}/2}+J_{\mathbf{k}+\mathbf{q}/2}+h & -\Delta\\
-\Delta & -\xi_{{\bf k}-\mathbf{q}/2}+J_{\mathbf{k}-\mathbf{q}/2}+h
\end{array}\right].
\end{equation}
By diagonalizing the matrix $\left[{\cal H}_{BdG}\right]$, we thus
obtain the two branches of the eigenvectors $[u_{{\bf k}\uparrow}^{(\eta)},v_{{\bf k}\downarrow}^{(\eta)}]^{T}$
and eigenvalues $\eta E_{{\bf k}\eta}$, labeled by the index $\eta=\pm$.
Explicitly, we have the single-particle excitation spectrum,
\begin{equation}
E_{\mathbf{k}\pm}=\sqrt{A_{\mathbf{k}}^{2}+\Delta^{2}}\pm B_{\mathbf{k}},
\end{equation}
where 
\begin{eqnarray}
A_{\mathbf{k}} & \equiv & \xi_{{\bf k}}+\frac{\hbar^{2}q^{2}}{8m}+\lambda\frac{\hbar^{2}qk_{y}}{4m},\\
B_{\mathbf{k}} & \equiv & \frac{\hbar^{2}qk_{x}}{2m}+J_{\mathbf{k}}+h.
\end{eqnarray}
Here, as usual we have reversed the sign for the eigenvalue $-E_{\mathbf{k}-}$,
in order to write the corresponding field operators of Bogoliubov
quasiparticles $\alpha_{\mathbf{k}-}$ and $\alpha_{\mathbf{k}-}^{\dagger}$
in the normal order. This introduces another energy shift $\mathcal{E}_{0}''=-\sum_{\mathbf{k}}E_{\mathbf{k}-}$
in the mean-field Hamiltonian. In the absence of altermagnetism or
external magnetic field, our expression of the single-particle excitation
spectrum recovers the results in Ref. \citep{Sheehy2015} and Ref.
\citep{Hong2025}, respectively. In the end, we obtain the diagonalized
mean-field Hamiltonian,
\begin{equation}
{\cal \mathscr{H}}_{MF}=\sum_{{\bf k}\eta=\pm}E_{{\bf k}\eta}\alpha_{{\bf k\eta}}^{\dagger}\alpha_{{\bf k}\eta}-\frac{\Delta^{2}}{U_{0}}\mathcal{S}+\mathcal{E}_{0},\label{mfhami}
\end{equation}
where 
\begin{equation}
\mathcal{E}_{0}=\mathcal{E}_{0}'+\mathcal{E}_{0}''=\sum_{\mathbf{k}}\left(A_{\mathbf{k}}-\sqrt{A_{\mathbf{k}}^{2}+\Delta^{2}}\right).
\end{equation}

\subsection{Thermodynamic potential}

From the diagonalized Hamiltonian, we can straightforwardly write
down the grand thermodynamic potential,
\begin{equation}
\Omega=-\frac{\Delta^{2}}{U_{0}}\mathcal{S}+\mathcal{E}_{0}-k_{B}T\sum_{{\bf k}\eta=\pm}\ln\left(1+e^{-\frac{E_{{\bf k\eta}}}{k_{B}T}}\right),
\end{equation}
where the last term is the standard contribution of non-interacting
fermionic particles to the thermodynamic potential. For given chemical
potential $\mu$ and temperature $T$, we have two independent parameters:
the order parameter $\Delta$ and the FF momentum $q$. These two
parameters should be determined by minimizing the grand thermodynamic
potential, i.e., 
\begin{eqnarray}
\frac{\partial\Omega}{\partial\Delta} & = & 0,\\
\frac{\partial\Omega}{\partial q} & = & 0.
\end{eqnarray}
Moreover, we need to satisfy the number equation, 
\begin{equation}
n=\frac{k_{F}^{2}}{2\pi}=-\frac{\partial\Omega}{\partial\mu},
\end{equation}
where $k_{F}$ is the Fermi wavevector.

\section{Numerical calculations}

In our numerical calculations, we typically take the Fermi wavevector
$k_{F}$ and Fermi energy $\varepsilon_{F}$ as the units for wavevector
and energy, respectively. This means that we can directly set $2m=\hbar=k_{B}=k_{F}=\varepsilon_{F}=1$.
In the dimensionless form, we find that the grand thermodynamic potential
is given by ($k_{x}=k\cos\varphi$ and $k_{y}=k\sin\varphi$),\begin{widetext}
\begin{equation}
\frac{\Omega}{\mathcal{S}k_{F}^{2}\varepsilon_{F}}=\frac{1}{4\pi^{2}}\int\limits _{0}^{2\pi}d\varphi\left[\int\limits _{0}^{k_{c}}dkC(k,\varphi)+\frac{c_{1}(\varphi)}{k_{c}}+\frac{c_{2}(\varphi)}{2k_{c}^{2}}+\cdots\right],
\end{equation}
where 
\begin{equation}
C(k,\varphi)=k\left[\frac{\Delta^{2}}{2k^{2}+\varepsilon_{B}}+A_{\mathbf{k}}-\sqrt{A_{\mathbf{k}}^{2}+\Delta^{2}}-T\sum_{\eta=\pm}\ln\left(1+e^{-\frac{E_{{\bf k}\eta}}{T}}\right)\right]
\end{equation}
and
\begin{eqnarray}
A_{\mathbf{k}} & = & k^{2}-\mu+\frac{q^{2}}{4}+\frac{\lambda qk\sin\varphi}{2},\\
E_{\mathbf{k}\pm} & = & \sqrt{A_{\mathbf{k}}^{2}+\Delta^{2}}\pm\left(qk\cos\varphi+\frac{\lambda k^{2}\sin2\varphi}{2}+h\right).
\end{eqnarray}
\end{widetext}Here, we have replaced the bare interaction strength
$U_{0}$ by the binding energy $\varepsilon_{B}$,
\begin{equation}
-\frac{\Delta^{2}}{U_{0}}=\frac{1}{4\pi^{2}}\int\limits _{0}^{2\pi}d\varphi\int\limits _{0}^{\infty}dkk\left[\frac{\Delta^{2}}{2k^{2}+\varepsilon_{B}}\right].
\end{equation}
Furthermore, to accurately calculate the integral over $C(k,\varphi)$,
we have introduced a high-momentum cut-off, i.e., $k_{c}\sim10k_{F}$.
Above $k_{c}$, we numerically expand, 
\begin{equation}
C(k,\varphi)=\frac{c_{1}(\varphi)}{k^{2}}+\frac{c_{2}(\varphi)}{k^{3}}+\cdots.
\end{equation}
The two coefficients $c_{1}(\varphi)$ and $c_{2}(\varphi)$ can be
found numerically by a linear curve fitting. We first calculate
\begin{eqnarray}
y_{1} & = & k_{1}^{3}C(k_{1},\varphi),\\
y_{2} & = & k_{2}^{3}C(k_{2},\varphi),
\end{eqnarray}
where, for example, $k_{1}=k_{c}\gg1$ and $k_{2}=2k_{c}\gg1$. Then,
we obtain, 
\begin{eqnarray}
c_{1}\left(\varphi\right) & = & \frac{y_{1}-y_{2}}{k_{1}-k_{2}},\\
c_{2}\left(\varphi\right) & = & \frac{y_{2}k_{1}-y_{1}k_{2}}{k_{1}-k_{2}}.
\end{eqnarray}

\subsection{Numerical procedure}

To find all the possible phases, for a given set of parameters, including
altermagnetic coupling $\lambda$, Zeeman field $h/E_{F}$, interaction
parameter $\varepsilon_{B}/\varepsilon_{F}$, and temperature $T/(\varepsilon_{F}/k_{B})$,
we minimize the thermodynamic potential against $\Delta$ and $q$.
As a simple and quick method, we shall use the Newton's gradient approach.
The various derivatives can be approximated by numerical differences.
This approach is efficient, provided a good guess of the initial parameters
for $\Delta$ and $q$, which may be obtained by looking at the contour
plot of the grand thermodynamic potential.

We try three different phases: the normal gas, BCS state and FF state.
Once these phases are determined, we calculate their free energy per
particle, $F/N=\Omega/N+\mu$, where $N=n\mathcal{S}$ is the total
number of particles. By comparing the free energy, we finally determine
which one is the ground state and then construct the whole phase diagram.

\section{Results and discussions}

In our numerical calculations, we always include a negligible temperature
$T=0.01T_{F}$, to soften the sharp Fermi surface at zero temperature
and hence to improve the numerical accuracy. We take a typical binding
energy $\varepsilon_{B}=0.08\varepsilon_{F}$, which leads to a BCS
pairing gap $\Delta_{0}=\sqrt{2\varepsilon_{B}\varepsilon_{F}}=0.4\varepsilon_{F}$
and a chemical potential $\mu_{0}=\varepsilon_{F}-\varepsilon_{B}/2=0.96\varepsilon_{F}$
in the absence of altermagnetism and magnetic field \citep{He2015}. 

\begin{figure*}
\begin{centering}
\includegraphics[width=0.45\textwidth]{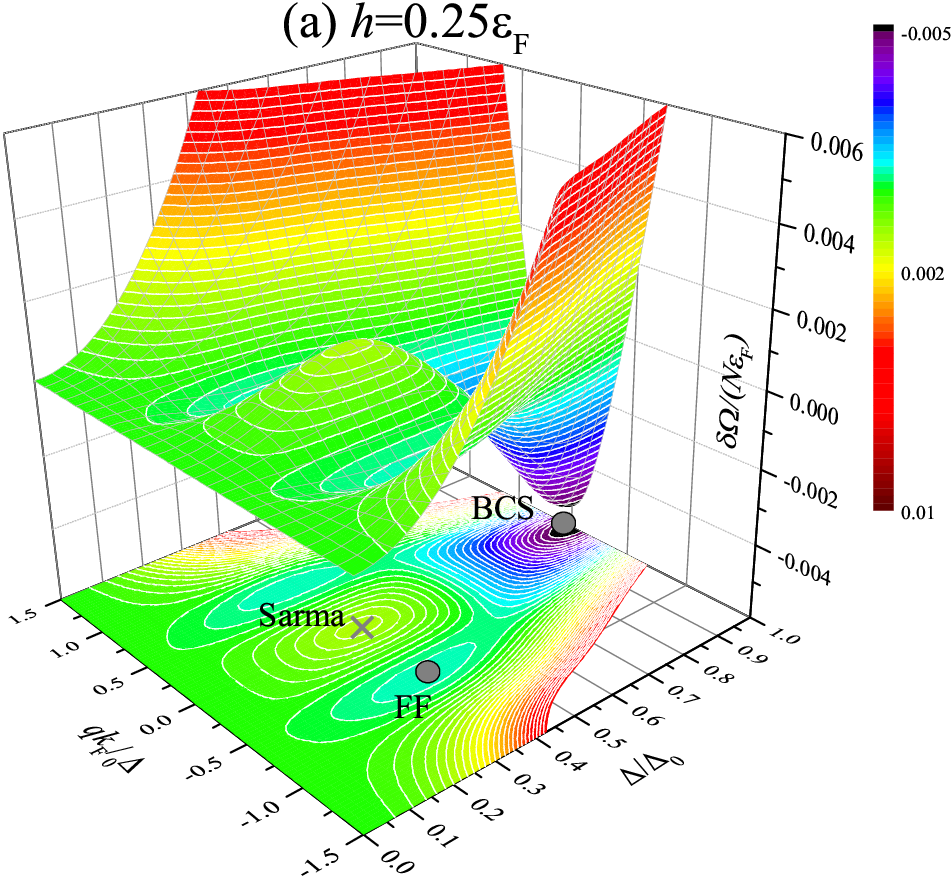}\includegraphics[width=0.45\textwidth]{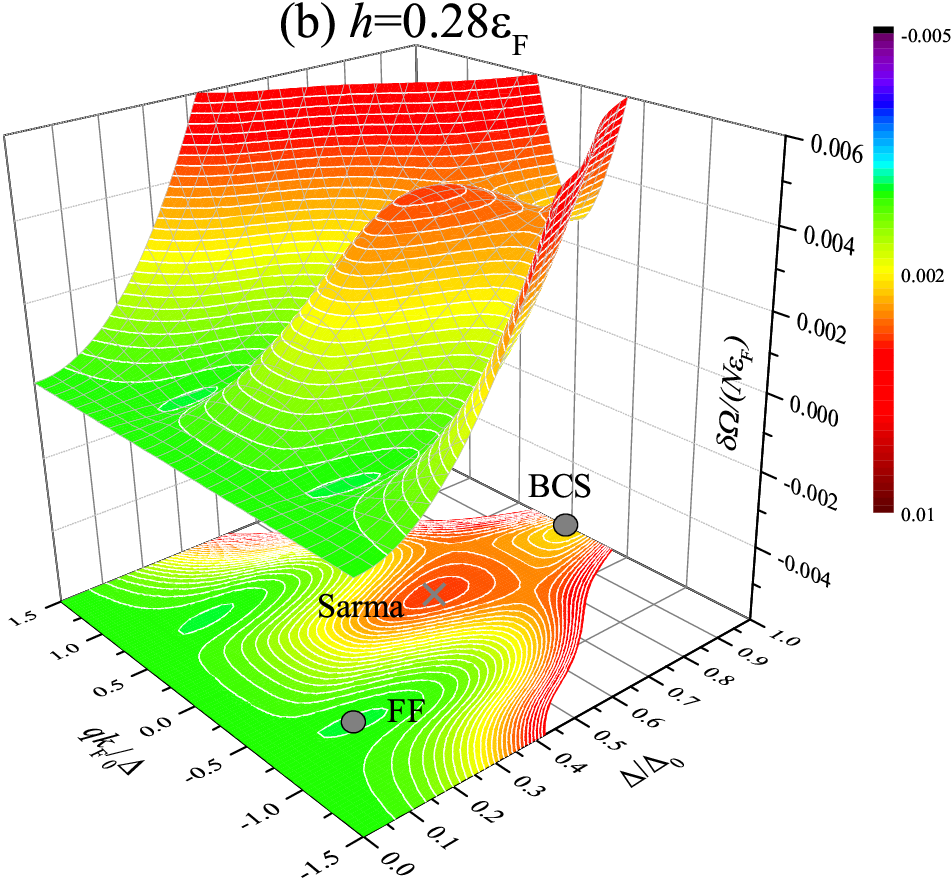}
\par\end{centering}
\caption{\label{fig2: Omega3DFFh} The landscape of the thermodynamic potential
$(\Omega-\Omega_{0})/(N\varepsilon_{F})$, as a function of the pairing
amplitude $\Delta$ and the FF momentum $q$, at two magnetic fields:
$h=0.25\varepsilon_{F}$ (a) and $h=0.28\varepsilon_{F}$ (b). Here,
$\Omega_{0}$ is the thermodynamic potential of the reference normal
state and we denote $\delta\Omega\equiv\Omega-\Omega_{0}$. The different
candidate phases are highlighted in the contour plot shown at the
bottom. We take a small altermagnetic coupling constant $\lambda=0.2$
and fix the chemical potential $\mu=\mu_{0}=0.96\varepsilon_{F}$.
The pairing amplitude $\Delta$ and the FF momentum are measured in
units of $\Delta_{0}$ and $\Delta_{0}/k_{F}$, respectively, where
$\Delta_{0}=0.4\varepsilon_{F}$ at the binding energy $\varepsilon_{B}=0.08\varepsilon_{F}$.}
\end{figure*}

\subsection{The FF state driven by magnetic field}

We first consider the conventional case, where the inhomogeneous FFLO
state is driven by magnetic field. In the absence of altermagnetism
(i.e., $\lambda=0$), for our chosen binding energy the first-order
phase transition from the BCS state to the FF state at zero temperature
is analytically known to occur \citep{Sheehy2015} at $h_{FF}=(\varepsilon_{B}/2)\sqrt{1+4\mu_{0}/\varepsilon_{B}}=0.28\varepsilon_{F}$.
The next transition from the FF state to a normal gas occurs \citep{Sheehy2015}
at $h_{c}=\sqrt{2\mu\varepsilon_{B}-\varepsilon_{B}^{2}}\sim0.39\varepsilon_{F}$.

In Fig. \ref{fig2: Omega3DFFh}, we show the landscape of the thermodynamic
potential $(\Omega-\Omega_{0})/(N\varepsilon_{F})$ in the $q-\Delta$
plane at two magnetic fields and in the presence of a small altermagnetic
coupling constant $\lambda=0.2$. We have subtracted the non-interacting
thermodynamic potential $\Omega_{0}$, so the preference of the BCS
state and the FF state over the normal state should be evident. Typically,
we find three candidate phases \citep{Hu2006}, two are local minima
(i.e., BCS at $q=0$ and FF at $q\neq0$) and one is a local maximum
(i.e., Sarma at $q=0$). Here, it is worth noting that the landscape
is symmetric with respect to $q$ (or more precisely $q_{x}$ in our
choice of the FF momentum) and we only need to consider the FF state
located at $q>0$.

\begin{figure}
\begin{centering}
\includegraphics[width=0.5\textwidth]{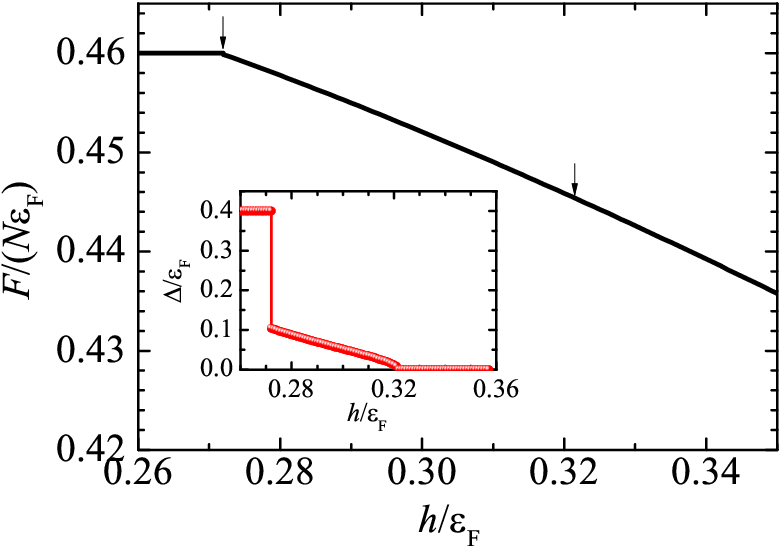}
\par\end{centering}
\caption{\label{fig3: EnergyL020} The free energy $F/(N\varepsilon_{F}$)
as a function of magnetic field, at a small altermagnetic coupling
constant $\lambda=0.2$. The two arrows indicate the transitions from
the BCS state to the FF$_{h}$ state and finally to the normal state.
The inset shows the pairing gap $\Delta$ as a function of the magnetic
field.}
\end{figure}

At the magnetic field $h=0.25\varepsilon_{F}$ (see Fig. \ref{fig2: Omega3DFFh}(a)),
the BCS state is clearly the absolute ground state. By increasing
magnetic field above the critical field $h_{FF}$, whose value is
slightly smaller than $0.28\varepsilon_{F}$ due to the existence
of the small altermagnetic coupling $\lambda=0.2$, the FF state becomes
more favorable, as illustrated in Fig. \ref{fig2: Omega3DFFh}(b).
The transition from the BCS state to the FF state should be first
order, as the two corresponding local minima are well separated by
an energy barrier (see, i.e., the saddle point near the local Sarma
maximum).

\begin{figure*}
\begin{centering}
\includegraphics[width=0.45\textwidth]{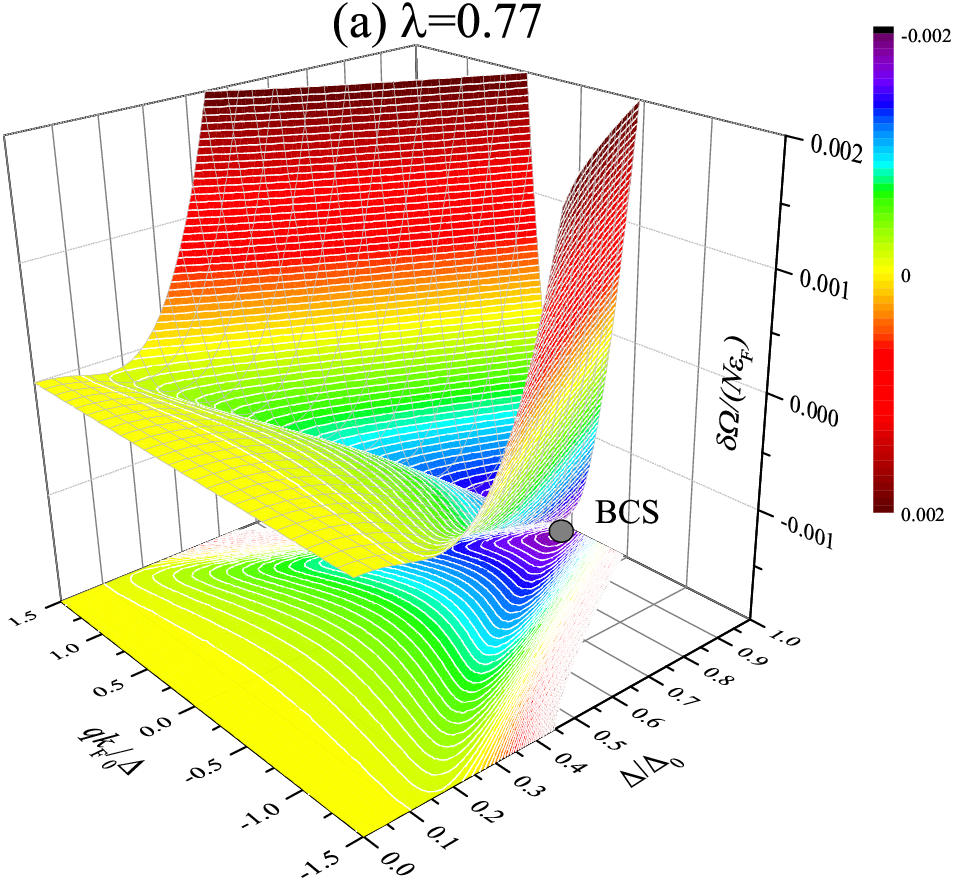}\includegraphics[width=0.45\textwidth]{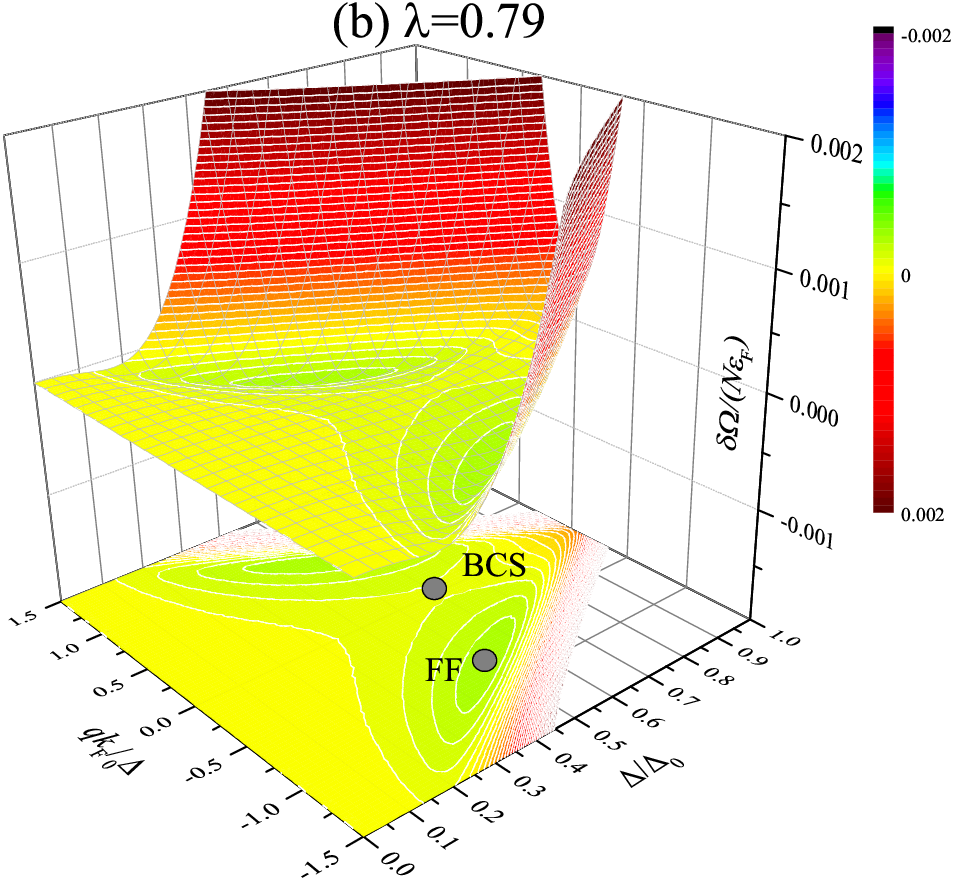}
\par\end{centering}
\begin{centering}
\includegraphics[width=0.6\textwidth]{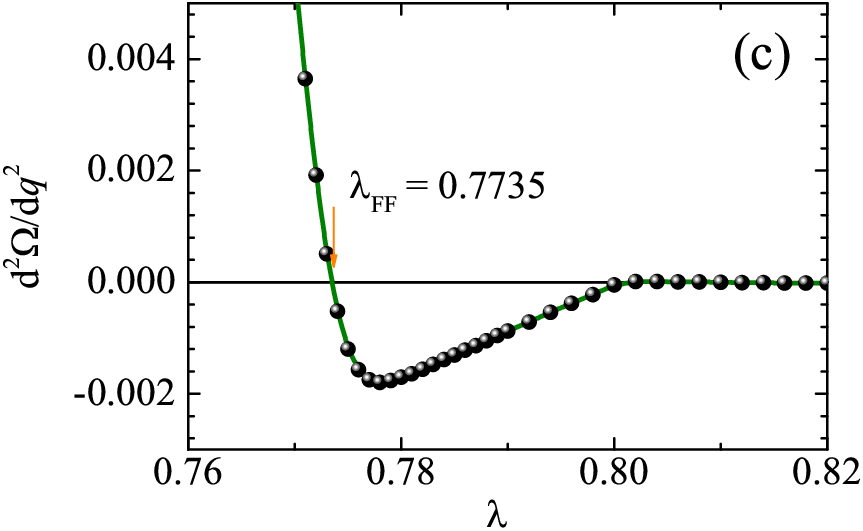}
\par\end{centering}
\caption{\label{fig4: Omega3DFFL} (a) and (b) The landscape of the thermodynamic
potential $(\Omega-\Omega_{0})/(N\varepsilon_{F})$, as a function
of the pairing amplitude $\Delta$ and the FF momentum $q$, at two
altermagnetic coupling constants: $\lambda=0.77$ and $\lambda=0.79$,
in the absence of magnetic field ($h=0$). At the bottom of the two
subfigures, we explicitly show the BCS and FF phases in the contour
plot. We have fixed the chemical potential $\mu=0.92831\varepsilon_{F}$.
(c) The instability of the BCS solution towards the FFLO state, as
signaled by the condition $\partial\Omega^{2}/\partial q^{2}<0$.
The quantity $\partial\Omega^{2}/\partial q^{2}$ is shown in units
of $\mathcal{S}\varepsilon_{F}^{2}$. The FFLO instability sets in
at $\lambda>\lambda_{FF}\simeq0.7735$.}
\end{figure*}

In Fig. \ref{fig3: EnergyL020}, we report the free energy per particle
as a function of magnetic field, in the presence of the small altermagnetic
coupling constant $\lambda=0.2$. The free energy exhibits a kink
at $h_{FF}\simeq0.272\varepsilon_{F}$, as a result of the first-order
phase transition from the BCS state to the FF state. This is in line
with a sudden decrease in the pairing gap $\Delta$, as shown in the
inset. By further increasing magnetic field, the pairing gap eventually
vanishes at $h_{c}\simeq0.322\varepsilon_{F}$, giving a smooth second-order
transition to the normal state.

We repeat the calculation of the free energy for other values of the
altermagnetic coupling constant $\lambda$ and determine how the parameter
window of the FF$_{h}$ phase changes. Here, the subscript ``$h$''
is used to indicate that the FF state is driven by magnetic field.
As shown in the phase diagram Fig. \ref{fig1: phasediagram} (see
the yellow area), the FF$_{h}$ phase window shrinks with increasing
$\lambda$. Eventually, the first-order transition line and the second-order
transition line meet at a quantum tri-critical Lifshitz point, located
at about $(\lambda,h)=(0.57,0.292\varepsilon_{F})$. A similar tri-critical
point for the FFLO phase is known to appear in a classical way by
increasing temperature \citep{Casalbuoni2004}. 

\subsection{The FF state driven by altermagnetism}

We now turn to consider the appearance of the FF state at the large
altermagnetic coupling. In Fig. \ref{fig4: Omega3DFFL}, we present
the landscape of the thermodynamic potential $(\Omega-\Omega_{0})/(N\varepsilon_{F})$
at the two altermagnetic coupling constants $\lambda=0.77$ and $\lambda=0.79$,
without magnetic field. At $\lambda=0.77$ (see Fig. \ref{fig4: Omega3DFFL}(a)),
we find only a global minimum at zero momentum with a pairing gap
$\Delta=\Delta_{0}$, corresponding to the standard BCS state. By
increasing $\lambda$ from $0.77$ to $0.79$, it turns out that such
as global minimum is rapidly lifted up and turns into a saddle point
with the pairing gap $\Delta<\Delta_{0}$. As illustrated in Fig.
\ref{fig4: Omega3DFFL}(b), such a saddle point, which we still refer
to as a BCS solution, suffers from an instability towards a nearby
minimum at $q\neq0$. This minimum is precisely the FF state driven
by altermagnetism, found earlier by Hong, Park and Kim in Ref. \citep{Hong2025},
although there is no net magnetic field in the $d$-wave altermagnetic
term.

In sharp contrast to the case of a large magnetic field, we observe
that a large altermagnetism does not introduce a local maximum in
the landscape of the thermodynamic potential, which acts as an energy
barrier for competing phases. Moreover, the FF state arises only after
the BCS solution gradually changes into a saddle point, with $\partial\Omega^{2}/\partial q^{2}<0$.
Therefore, the phase transition from the BCS state to the FF state
is smooth. To confirm this observation, in Fig. \ref{fig4: Omega3DFFL}(c)
we plot the quantity $\partial\Omega^{2}/\partial q^{2}$ with increasing
altermagnetic coupling constant $\lambda$. Indeed, we find that $\partial\Omega^{2}/\partial q^{2}$
changes its sign at a critical altermagnetic coupling $\lambda_{FF}=0.773$.
Beyond this critical value, the FF state becomes the absolute ground
state.

\begin{figure}
\begin{centering}
\includegraphics[width=0.5\textwidth]{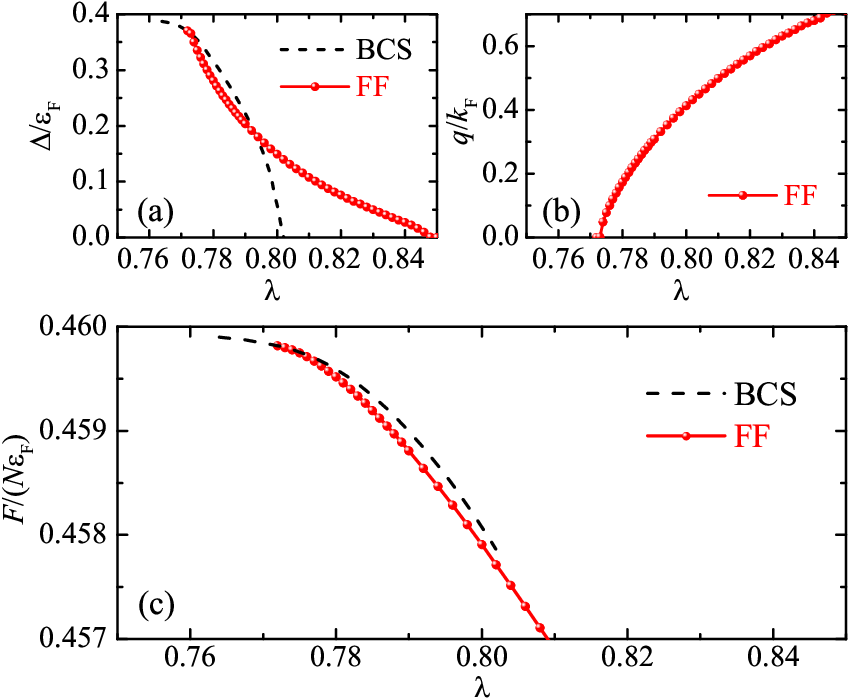}
\par\end{centering}
\caption{\label{fig5: Energyh0} The pairing gap $\Delta/\Delta_{0}$, the
FF momentum $q/k_{F}$, and the free energy $F/(N\varepsilon_{F}$)
of the BCS state and the FFLO state, as a function of the altermagnetic
coupling constant, in the absence of magnetic field ($h=0$).}
\end{figure}

The above instability analysis is useful to accurately determine the
critical altermagnetic coupling $\lambda_{FF}$. However, it does
not help to locate another phase transition point $\lambda_{c}$ from
the FF state to the normal state, since the metastable BCS solution
ceases to exist once $\lambda>0.8$. To determine the critical altermagnetic
coupling $\lambda_{c}$, we calculate the pairing gap and the free
energy of the altermagnetism-driven FF state and compare them with
the corresponding BCS results in Fig. \ref{fig5: Energyh0}(a) and
Fig. \ref{fig5: Energyh0}(c), respectively.

As expected, as a result of the second-order phase transition, the
FF pairing gap smoothly connect to the BCS pairing gap at $\lambda_{FF}$.
The two free energies also smooth connect. Furthermore, the FF momentum
becomes nonzero precisely at $\lambda_{FF}$ (see Fig. \ref{fig5: Energyh0}(b))
\citep{Hong2025}. On the other hand, the value of the critical altermagnetic
coupling $\lambda_{c}\simeq0.848$ can be extracted by tracing the
FF pairing gap to vanish. Therefore, without magnetic field ($h=0$),
the altermagnetic metal with a two-body binding energy $\varepsilon_{B}=0.08\varepsilon_{F}$
is an inhomogeneous FF superconductor in a relatively narrow region,
$0.773<\lambda<0.848$. 

\begin{figure}
\begin{centering}
\includegraphics[width=0.5\textwidth]{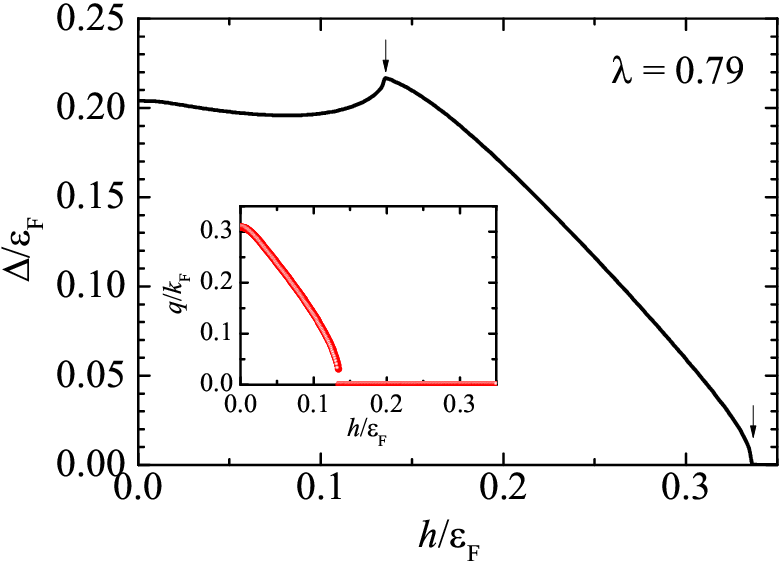}
\par\end{centering}
\caption{\label{fig6: DeltaQL079} The pairing gap $\Delta$ (main figure)
and the FF momentum $q$ (inset), as a function of the magnetic field,
at a large altermagnetic coupling constant $\lambda=0.79$. The two
arrows indicate the second-order phase transitions from the FF$_{\lambda}$
state to the polarized BCS state and finally to the normal state.}
\end{figure}

To investigate how the altermagnetism-driven FF state is affected
by magnetic field, in Fig. \ref{fig6: DeltaQL079} we present the
pairing gap (main figure) and the FF momentum (inset) of the ground
state, as a function of magnetic field. At zero magnetic field ($h=0$),
the Fermi system is initially in the FF state with a pairing gap $\Delta\simeq0.2\varepsilon_{F}$
and a FF momentum $q\simeq0.3k_{F}$. By increasing magnetic field,
we find the pairing gap of the FF ground state shows a non-monotonic
dependence, up to a critical field $h_{FF}\simeq0.135\varepsilon_{F}$.
Beyond this critical value, the system turns into a BCS state. This
smooth transition can be double checked by looking at the FF momentum
$q$ in the inset, which precisely vanishes at $h_{FF}$. By furthering
increase the magnetic field, we find the pairing gap of the BCS state
eventually becomes zero at the second critical field $h_{c}\simeq0.336\varepsilon_{F}$
and the system becomes normal.

We perform the calculations at different altermagnetic couplings in
the interval $0.773<\lambda<0.8$ and observe the similar magnetic
field dependence. At another interval $0.8<\lambda<0.848$, the Fermi
system turns out to directly transit from the FF state to the normal
state at the critical field $h_{c}$. In the phase diagram Fig. \ref{fig1: phasediagram},
we highlight the parameter window of the FF$_{\lambda}$ phase with
blue color, where the subscript ``$\lambda$'' now represents the
altermagnetism-driven FF state. The window shrinks with increasing
magnetic field and ends at a non-trivial quantum critical point $(\lambda,h)\simeq(0.80,0.285\varepsilon_{F})$,
where the two second-order phase transition curves intersect (see
the purple dot in the phase diagram).

\begin{figure}
\begin{centering}
\includegraphics[width=0.5\textwidth]{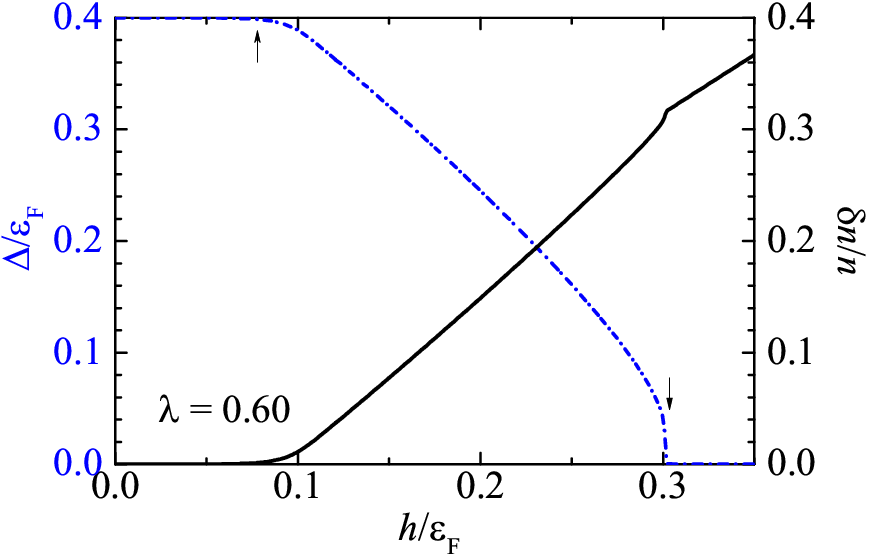}
\par\end{centering}
\caption{\label{fig7: pBCSdnL060} The pairing gap $\Delta$ (left axis) and
the spin-population imbalance $\delta n$ (right axis), as a function
of the magnetic field, at the altermagnetic coupling constant $\lambda=0.60$.
The two arrows indicate the crossover from the BCS state to the polarized
BCS state and the second-order phase transition from the polarized
BCS state to the normal state, respectively.}
\end{figure}

\subsection{The polarized BCS state}

As the spin-population imbalance $\delta n=n_{\uparrow}-n_{\downarrow}$
of a FF state is nonzero, the continuous phase transition from the
FF state to the BCS state shown in Fig. \ref{fig6: DeltaQL079} at
the critical field $h_{FF}\simeq0.135\varepsilon_{F}$ is highly non-trivial.
It is suggestive that the spin-population imbalance of the BCS state
is also nonzero. Hence, we obtain a partially spin-population polarized
BCS state as the ground state at zero temperature, which can hardly
be realized in other interacting Fermi systems \citep{Forbes2005,Zou2018}.

In fact, in the phase diagram Fig. \ref{fig1: phasediagram} we find
a large parameter window for the polarized BCS state. To be concrete,
in Fig. \ref{fig7: pBCSdnL060} we show the pairing gap $\Delta$
and the spin-population imbalance $\delta n$ of the ground state
at the altermagnetic coupling $\lambda=0.6$, with increasing magnetic
field. The pairing gap decreases from the constant value $\Delta_{0}$
at a critical field $h_{\textrm{pBCS}}\simeq0.076\varepsilon_{F}$.
At the same critical field, the spin-population imbalance $\delta n$
becomes nonzero and starts to increase with magnetic field almost
linearly.

\begin{figure}
\begin{centering}
\includegraphics[width=0.24\textwidth]{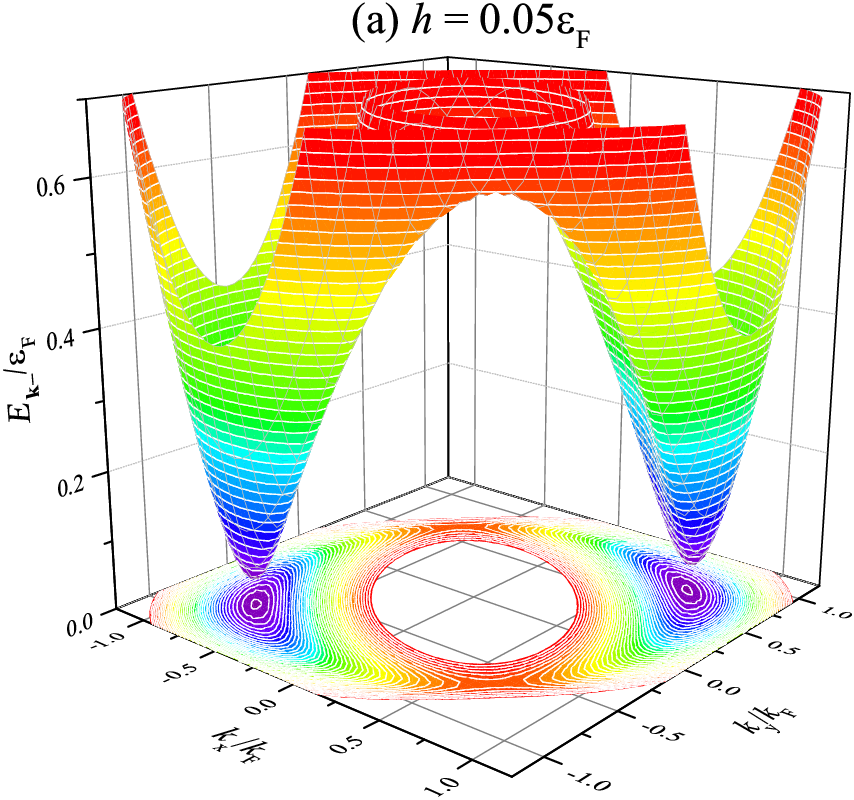}\includegraphics[width=0.24\textwidth]{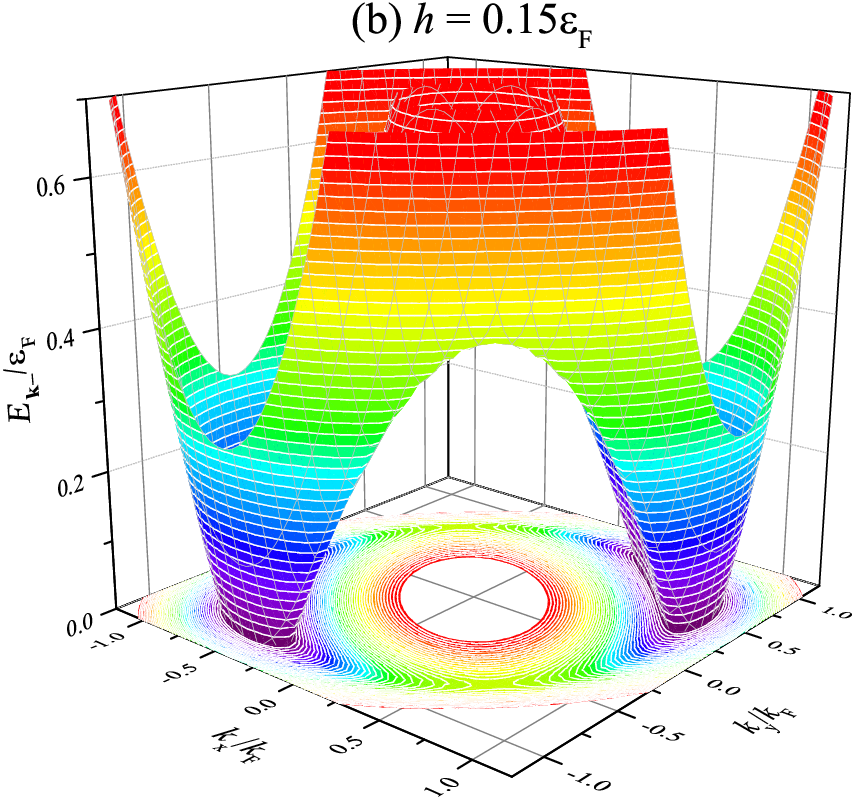}
\par\end{centering}
\caption{\label{fig8: EkmL060} The single-particle excitation spectrum $E_{\mathbf{k}-}$
of a BCS state at $h=0.05\varepsilon_{F}$ (a) and of a polarized
BCS state at $h=0.15\varepsilon_{F}$, in the presence of an altermagnetic
coupling $\lambda=0.60$. With increasing magnetic field, the excitation
spectrum becomes gapless and supports a nonzero spin-population imbalance.}
\end{figure}

At nearly zero temperature a nonzero spin-population imbalance can
only be supported, when the single-particle excitation spectrum of
the BCS state becomes gapless. To confirm this, in Fig. \ref{fig8: EkmL060}
we present the Bogoliubov spectrum $E_{\mathbf{k}-}$ at the two magnetic
fields $h=0.05\varepsilon_{F}$ (a) and $h=0.15\varepsilon_{F}$ (b).
In the former case, which is below the critical field $h_{\textrm{pBCS}}\simeq0.076\varepsilon_{F}$,
the spectrum is gapped, consistent with the fact that $\delta n\simeq0$.
In the latter case, we see clearly that the spectrum $E_{\mathbf{k}-}$
becomes gapless around the two diagonal points$(k_{x},k_{y})\sim[\pm k_{F}\cos(\pi/4),\pm k_{F}\sin(\pi/4)]$
in momentum space.

It is readily seen that we can use the gapless condition $E_{\mathbf{k}-}=0$
at $q=0$, or more explicitly, 
\begin{equation}
h=\sqrt{\left(k^{2}-\mu_{0}\right)^{2}+\Delta_{0}^{2}}-\frac{\lambda k^{2}\sin2\varphi}{2}
\end{equation}
to estimate the critical field $h_{\textrm{pBCS}}$ at a given altermagnetic
coupling constant $\lambda$. By minimizing the right-hand-side of
the equation with respect to $k^{2}$ and $\varphi$, we find that
\begin{equation}
h_{\textrm{pBCS}}=\frac{1}{2}\left[\sqrt{4-\lambda^{2}}\Delta_{0}-\lambda\mu_{0}\right].
\end{equation}
This estimation seems to be consistent with our numerical calculation
(i.e., see the dot-dashed line in Fig. \ref{fig1: phasediagram}).
In particular, by setting $h_{\textrm{pBCS}}=0$, we obtain a threshold
altermagnetic coupling $\lambda_{\textrm{thres}}=2\Delta_{0}/\sqrt{\Delta_{0}^{2}+\mu_{0}^{2}}\simeq0.77$.
This value agrees excellently well with the critical altermagnetic
coupling $\lambda_{FF}=0.773$ at zero magnetic field.

Finally, in the weak coupling regime ($\varepsilon_{B}\ll\varepsilon_{F}$)
that we are interested in, since the threshold coupling $\lambda_{\textrm{thres}}\simeq2\Delta_{0}/\varepsilon_{F}$
scales linearly with the BCS pairing gap $\Delta_{0}$, we anticipate
that our phase diagram Fig. \ref{fig1: phasediagram} qualitatively
holds for other binding energies, aside from an overall rescaling
in the energy scale.

\section{Conclusions and outlooks}

To conclude, we have investigated the inhomogeneous FFLO in an altermagnetic
metal, driven by either external magnetic field or $d$-wave altermagnetism.
We have confirmed the emergence of the FFLO state at large altermagnetic
coupling in the presence of an $s$-wave interparticle attraction
and thereby have clarified its formation mechanism, which is not reconciled
in the recent two studies \citep{Chakraborty2024,Hong2025}. We have
also clarified the nature of quantum phase transitions occurring between
competing phases and have constructed a universal rich phase diagram
Fig. \ref{fig1: phasediagram}.

Quite unexpectedly, we have found a uniform, partially spin-populated
polarized BCS state, due to the altermagnetic coupling term. This
polarized BCS phase has a significant parameter space in the phase
diagram. Due to its gapless single-particle excitation spectrum, we
anticipate that very intriguing collective excitations would arise
in the polarized BCS state. Its dynamic response and sound propagation
would be an interesting topic for future studies \citep{Zou2018,Hu2010}.

Furthermore, we have also identified the existence of two quantum
tri-critical Lifshitz points \citep{Hornreich1975,Gubbels2009,Pisarski2019}.
A brief discussion of the related Lifshitz physics based on Ginzburg-Landau
free-energy functional will be presented elsewhere \citep{Hu2025}.
\begin{acknowledgments}
This research was supported by the Australian Research Council's (ARC)
Discovery Program, Grants Nos. DP240101590 (H.H.) and DP240100248
(X.-J.L.). 
\end{acknowledgments}

\end{document}